\documentclass{amsart}

\usepackage{amsmath, amssymb, amsthm}
\usepackage[colorlinks=true, citecolor=red, linkcolor=red, urlcolor=blue]{hyperref}
\usepackage{fontawesome}
\usepackage{graphicx}
\usepackage{csquotes}
\usepackage{subcaption}
\usepackage{booktabs}

\newcommand{\SSIM}{\ensuremath{\mathrm{SSIM}}}

\title{Enhanced Low-Dose CT Image Reconstruction by Domain and Task Shifting Gaussian Denoisers}

\author{Tim~Selig$^{*1}$ \href{mailto:tim.selig@h-da.de}{\faEnvelope} \and Thomas~März$^1$ \href{mailto:thomas.maerz@h-da.de}{\faEnvelope} \and Martin~Storath$^2$ \href{mailto:martin.storath@thws.de}{\faEnvelope} \and Andreas~Weinmann$^1$ \href{mailto:andreas.weinmann@h-da.de}{\faEnvelope}}

\thanks{\textnormal{T.~S., T.~M. and A.~W. acknowledge support by the Hessian Ministry of Higher Education, Research, Science, and the Arts through the \enquote{Programm zum Aufbau eines akademischen Mittelbaus an hessischen Hochschulen}.
M.~S. acknowledges support by the project DIBCO funded by the research program 
\enquote{Informations- und Kommunikationstechnik} of the Bavarian State Ministry of Economic Affairs, 
Regional Development and Energy  (DIK-2105-0044 / DIK0264).
A.~W. acknowledges support of Deutsche Forschungsgemeinschaft (DFG) under project number 514177753.}}
\thanks{\textnormal{${^{*}}$Corresponding author: Tim~Selig (tim.selig@h-da.de).}}

\begin{document}

% Title and footnotes
\maketitle

\vspace{-2em}
\begin{center}
    \small
    $^1$ Hochschule Darmstadt, ACIDA Lab, 64287 Darmstadt, Germany\\
    $^2$ Technische Hochschule W\"urzburg-Schweinfurt, Lab for Mathematical Methods in Computer Vision and Machine Learning, 97421 Schweinfurt, Germany
    \normalsize
\end{center}

% Abstract
\begin{abstract}
    Computed tomography from a low radiation dose (LDCT) is challenging due to high noise in the projection data. Popular approaches for LDCT image reconstruction are two-stage methods, typically consisting of the filtered backprojection (FBP) algorithm followed by a neural network for LDCT image enhancement. Two-stage methods are attractive for their simplicity and potential for computational efficiency, typically requiring only a single FBP and a neural network forward pass for inference. However, the best reconstruction quality is currently achieved by unrolled iterative methods (Learned Primal-Dual and ItNet), which are more complex and thus have a higher computational cost for training and inference. We propose a method combining the simplicity and efficiency of two-stage methods with state-of-the-art reconstruction quality. Our strategy utilizes a neural network pretrained for Gaussian noise removal from natural grayscale images, fine-tuned for LDCT image enhancement. We call this method FBP-DTSGD (Domain and Task Shifted Gaussian Denoisers) as the fine-tuning is a task shift from Gaussian denoising to enhancing LDCT images and a domain shift from natural grayscale to LDCT images. An ablation study with three different pretrained Gaussian denoisers indicates that the performance of FBP-DTSGD does not depend on a specific denoising architecture, suggesting future advancements in Gaussian denoising could benefit the method. The study also shows that pretraining on natural images enhances LDCT reconstruction quality, especially with limited training data. Notably, pretraining involves no additional cost, as existing pretrained models are used. The proposed method currently holds the top mean position in the LoDoPaB-CT challenge.
\end{abstract}

% Introduction
\section{Introduction}

\begin{figure}[htbp] 
    \centering
    \includegraphics[width=\textwidth]{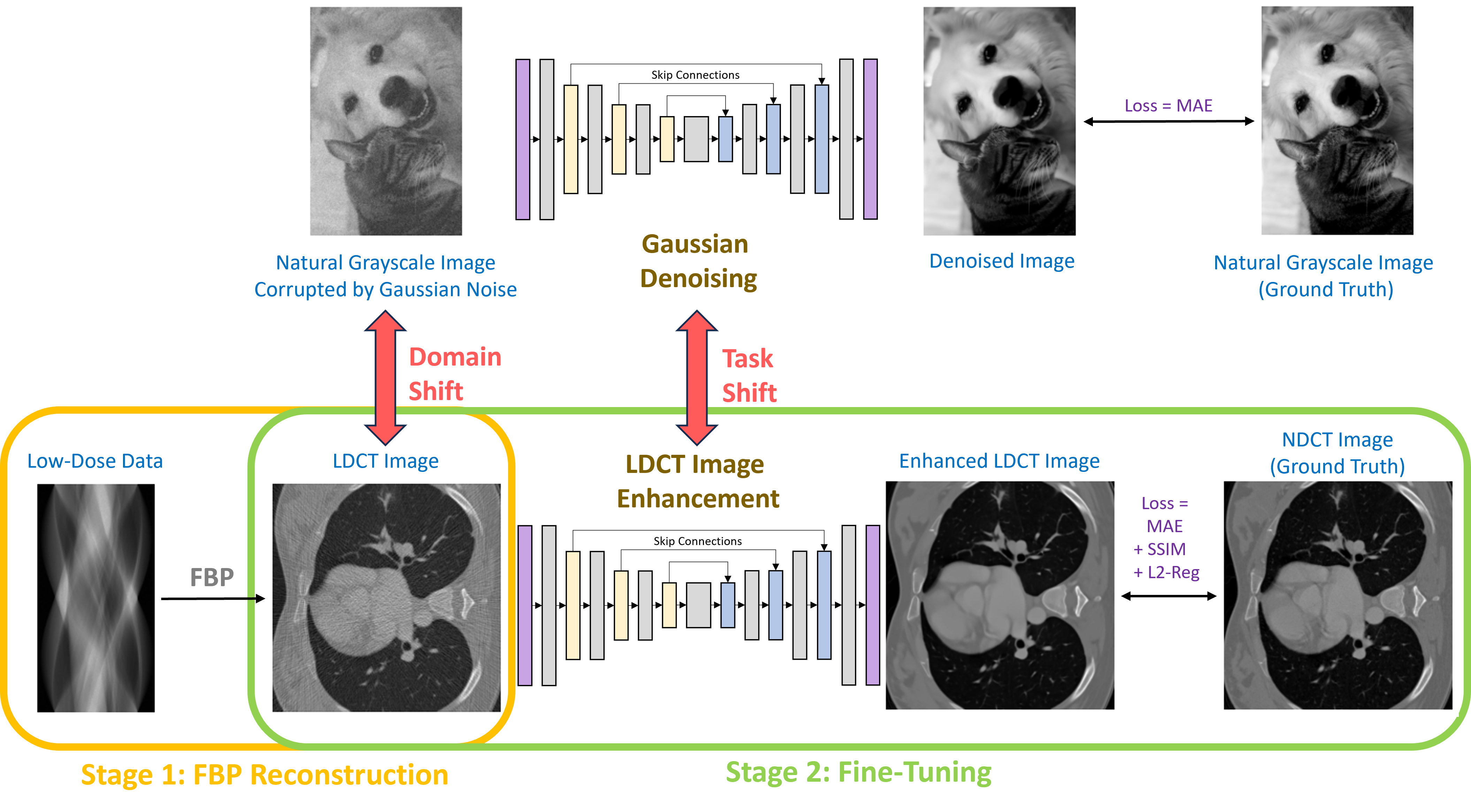}
    \caption{Overview of the proposed FBP-DTSGD methodology. The proposed approach consists of a two-stage process for LDCT image enhancement. 
    A FBP is applied to the LDCT data to obtain an initial reconstruction. 
    A pretrained Gaussian denoiser is fine-tuned for the downstream task, i.e., for the enhancement of CT-images obtained from the FBP-stage.
    }
\label{fig:method}
\end{figure}

Computed tomography (CT) is a cornerstone of modern medical imaging, providing accurate and non-invasive visualization of the internal structures of the human body. Its widespread use is due to its versatility, as it serves as a diagnostic tool for a wide range of medical examinations \cite{lell2015evoct, kalender2011ctfund}. 
Despite a long history of contributions, the reconstruction of CT images remains an active area of research due to the continuous search for higher image quality, less artifacts and reduced noise while at the same time reducing the ionizing radiation \cite{yang2018ldctgan}. 
In particular, reconstruction from low radiation dose projections 
and reconstruction from sparse or limited  projection data remain challenging tasks. 
This work is focused on low-dose CT (LDCT) reconstruction,
which is subject to intensive research, see e.g.~\cite{lu2023ldctreco, hu2023ultra, kulathilake2023review, li2023transformer, wang2023ctformer, mazandarani2023unext, xiong2023reunet}.

Until a few years ago, reconstructions from iterative variational regularization methods were among the state-of-the-art for CT image reconstruction; 
for example, these methods are based on total variation \cite{liu2012adaptive}, total generalized variation \cite{niu2014sparse} or 
Hessian Schatten norms \cite{liu2017low} regularization, to mention only a few.
In recent years, deep learning methods have led to a paradigm shift in medical image reconstruction. 
In particular, convolutional neural networks (CNNs) have demonstrated convincing
capabilities in mitigating artifacts 
and reconstruction noise \cite{litjens2017dlmedimg}.
One may distinguish between fully learned methods, iterative approaches, and multi-stage methods. 
Fully learned methods involve the direct reconstruction of images through deep learning architectures, utilizing raw projection data or sinograms as input. 
Iterative approaches, such as statistical iterative reconstruction or model-based iterative reconstruction (MBIR) aim at refining image quality \cite{stiller2018iterative}
by alternating between reconstruction and denoising.
Several studies employing MBIR techniques, including the Plug-and-Play framework~\cite{venkatakrishnan2013pnp} for the integration of deep learning-based denoisers, have consistently showcased substantial improvements in image quality, robustness to noise, and overall enhancement of the medical image reconstruction process~\cite{ye2018ctpnp, zhang2021pnp}.
Two-stage and Multi-stage methods employ neural networks as a step or as steps in a pipline to improve the results of an initial reconstruction. Typically, filtered backprojection (FBP) is utilized in the initial reconstruction phase. In a subsequent stage, or in subsequent stages, neural networks are employed to perform image enhancement tasks which improve the reconstruction quality. Successful approaches of this type in particular using a UNet-type architecture in the second stage are e.g.~\cite{zhou2018unetpp, mazandarani2023unext,xiong2023reunet,liu2020istaunet,leuschner2021lodopab,liu2022musc} (for further discussion we refer to the related work section).
Two-stage methods are attractive for their structural simplicity and potential for computational efficiency, 
as they typically require only a single FBP and a forward pass through a network during inference.

To assess the LDCT image reconstruction quality, the LoDoPaB-CT challenge has been established by Leuschner et al.~\cite{leuschner2021lodopabcomp}.
The public dataset consists of pairs of low-dose sinograms
and corresponding normal-dose CT (NDCT) images. 
A hidden challenge test set is used for the evaluation of the methods.

While two-stage methods are attractive for their simplicity and efficiency, unrolled iterative methods have achieved even higher reconstruction quality, as demonstrated by the top-ranking methods in the LoDoPaB-CT challenge:
The Learned Primal-Dual~\cite{adler2018learnedprimaldual} is based on unrolling a proximal primal-dual optimization method, but where
the proximal operators are replaced with CNNs.
ItNet~\cite{genzel2022nearexactrecovery}  is an unrolled iterative network that integrates FBP, 
a UNet trained on CT data, and repeatedly applies image enhancement and data consistency steps. 
While these methods achieve a higher reconstruction quality than two-stage methods, they
have a considerably more complex architecture 
and are computationally more demanding
as they involve multiple applications of the 
forward imaging operator and the FBP.

\subsection{Proposed method and contribution}
In this paper, we propose  a two-stage method 
that achieves state-of-the-art reconstruction quality.
As common in two-stage methods, the first stage is a FBP reconstruction, and the second stage is a neural network trained for LDCT image enhancement.
The key strategy is to employ a network designed and trained for Gaussian noise removal of natural grayscale images as the second stage.
This particular choice is motivated by the fact that Gaussian denoising of natural images is a well-studied task, and in particular, a series of pretrained models are available.

Such a pretrained Gaussian denoiser is then adapted
for the downstream task of enhancing LDCT images.
This effectively means a domain shift (from natural grayscale images to CT images) and a task shift (from Gaussian noise removal to LDCT image enhancement). 
Therefore, we refer to the proposed methodology as FBP-DTSGD (Domain and Task Shifted Gaussian Denoisers).
The shift can be easily implemented by just adapting the input and output dimensions of the data to align with the network architectures. 
As an addition, we augment the loss with an SSIM-based term.
These modifications preserve the core network architecture.

By default we use the pretrained Gaussian denoiser DRUNet of Zhang et al.~\cite{zhang2021pnp} for the second stage.
Fine-tuning this on the LoDoPaB-CT dataset, we achieve a top mean rank in the LoDoPaB-CT challenge~\cite{leuschner2021lodopabcomp}.
This shows that the proposed method is capable of achieving  state-of-the-art results for LDCT image reconstruction.

An ablation study is conducted on subsets of the LoDoPaB-CT dataset, which utilizes parallel-beam geometry and the 2016 Low Dose CT Grand Challenge (2016LDCTGC) dataset~\cite{mccollough2017low}, which employs fan-beam geometry in a helical scan process.
We compare three different architectures of pretrained Gaussian denoiser which have been pretrained on different sets of natural grayscale images.
Besides the DRUNet~\cite{zhang2021pnp}, we also test the KBNet~\cite{zhang2023kbnet}, which is a UNet-based architecture with a kernel basis attention module,
and the Restormer~\cite{zamir2022restormer} which is a transformer-based architecture.
The results indicate that the  proposed FBP-DTSGD methodology does not depend on a specific architecture of the Gaussian denoiser.
The study also shows that the pretraining of the Gaussian denoiser on natural images is beneficial, 
and that this benefit is most pronounced if only a small amount of training data is used.
 
We evenually highlight the main benefits of the proposed method: as two-stage process, it is computationally efficient, as it only requires a single FBP and a single forward pass through a network for inference.
Furthermore, the approach is architecturally simple, in particular, it does not require a specialized architecture for LDCT image enhancement, nor does it rely on applying CT-specific operators beyond the initial application of the standard FBP reconstruction.
Eventually, this work shows that
Gaussian denoisers pretrained on natural images 
are suitable for being shifted to an image enhancement 
on the LDCT  domain. So future improvements in Gaussian denoising may directly benefit the proposed method.

Preliminary results for the proposed method were presented at the 50th International Conference on Applications of Mathematics in Engineering and Economics 2024~\cite{selig2024fine}.
The present manuscript substantially extends upon the proceeding 
by providing a thorough discussion of related work and a significantly extended experimental analysis and evaluation.
In particular, additional Gaussian denoiser architectures (KBNet and the transformer-based UNet-type model Restormer) pretrained on different datasets are investigated.
Further, experiments were performed on an additional dataset (2016 Low-Dose CT Grand Challenge).
Moreover, the inference times were analyzed and compared to ItNet, the previously top-ranked method in the LoDoPaB-CT challenge.
Finally, the methodology was tested on the full LoDoPaB-CT dataset, with a detailed analysis of the effectiveness of using pretrained weights.

\subsection{Related Work}

The original \emph{UNet} was introduced by Ronneberger et al. \cite{ronneberger2015unet},
which found widespread utilization for medical image processing. 
However, its applications extend beyond the medical domain. In non-medical scenarios, UNets have proven effective, as demonstrated in works such as \cite{zhang2018road, wu2023uiunet, jansson2017singing}.
UNet++, developed by Zhou et al. \cite{zhou2018unetpp}, enhances the capabilities of the original UNet.
While the original UNet employs a basic architecture with skip connections, UNet++ introduces nested and dense convolution blocks to address the limitation of combining semantically dissimilar feature maps. 
UNet++ aims to progressively capture fine-grained details through these innovations. 
Furthermore, variations in loss functions, optimization techniques, and training strategies across the networks are described in \cite{zhou2018unetpp}.
Liu et al. introduce the ISTA UNet \cite{liu2020istaunet}, an adaptation based on the encoder-decoder structure of the original UNet, that leverages task-driven dictionary learning and sparse coding. Unlike the original UNet, ISTA UNet replaces the encoder with a sparse representation and linearizes the decoder, showcasing versatility across datasets with sensitivity to hyperparameter choices.
Liu et al. introduce a multiscale UNet-like sparse coding (MUSC) approach in \cite{liu2022musc},
exploring the development of multiscale convolutional dictionaries. 
Xiong et al. introduce Re-UNet, a novel reverse U-shaped network design \cite{xiong2023reunet}, motivated by enhancing LDCT image quality. Re-UNet employs a upsampling-downsampling approach, aiming to preserve texture details in medical CT images.
Additionally, Mazandarani et al. contribute to LDCT denoising with their UNeXt architecture, 
integrating modified ConvNeXt blocks into a UNet model~\cite{mazandarani2023unext}.

In addition to UNet-based models and their variants, there are various alternative deep learning-based approaches.  
The integration of encoder-decoder architectures with Generative Adversarial Networks (GANs) 
contributed to advancements in reconstruction quality and artifact reduction~\cite{chen2017encdec, yang2018ldctgan, kim2020ldctgan, wolterink2017gan, shan2018encdec, hu2019wgan, yi2018gan, fan2020qae, ma2020gan}.
Beyond these methods, there are reconstruction strategies based on wavelet-based methods as well as transformer-based models, to mention only two classes; for overview and discussion including other novel approaches we refer to~\cite{pelt2018msdcnn, wang2023ctformer, liang2020edcnn, denker2020ldct, kang2016wavelets, gho2019ldct, li2023transformer}.

Moreover, parameter-efficient pretraining and adaptation methods, such as Low-Rank Adaptation (LoRA), aim to reduce the number of trainable parameters by injecting low-rank decomposition matrices into each layer of a pretrained model. Originally developed for transformers~\cite{hu2021lora}, LoRA has more recently been explored for CNNs in the context of fine-tuning models with fewer parameters, known as LoRA-C~\cite{ding2024lorac}.

In the early days of deep learning, models were typically trained from scratch. However, recent advances in computer vision and natural language processing have shown that fine-tuning pretrained models can significantly improve performance~\cite{radford2018improving, liu2022swin}. Fine-tuning involves updating all trainable parameters of a pretrained model to adapt it to a specific task~\cite{zheng2023learn}, which we refer to as a task shift. In some cases, this adaptation may involve a domain shift, where the data distributions of the source and target domains differ -- a process referred to as domain adaptation~\cite{farahani2021domainadap, yuqi2024domainadap}.
Our approach uses models that are pretrained on large-scale datasets, typically containing thousands of images, which are then fine-tuned for a different domain. This fine-tuning is done on datasets of different sizes, including cases with very limited training data. When the training dataset is particularly small, the process can be referred to as few-shot learning, which aims to generalize from very limited training data ~\cite{song2022fsl}.
It is discussed in the literature that fine-tuning can have positive benefits, but that the strategies must be chosen properly, and the problems must be amenable, e.g., \cite{hendrycks2019pretraining, he2019pretraining, xu2023parameter, lin2024pretraining}.
Zheng et al.~\cite{zheng2023learn} focuses in their survey on utilizing and adapting Foundation Models (FM) for various downstream tasks. 
The paper categorizes so called Learn From Model techniques into five main areas: 
model tuning, model distillation, model reuse, meta-learning, and model editing.
Model distillation is about compressing models to reduce computational complexity while maintaining performance~\cite{gou2021knowledge}.
Model reuse combines predictions from multiple models to improve overall performance~\cite{jiang2023llm}.
Meta-learning, or \enquote{learning to learn from models}, enables models to quickly adapt to new tasks~\cite{hospedales2022metalearning}, 
and model editing allows for updating model knowledge without affecting performance on other inputs~\cite{mitchell2022fast}.
The area in which our proposed method is located is the area of model tuning, which inherits many subcategories.
Zheng et al.~\cite{zheng2023learn} focuses primarily on Fine-Tuning, Adapter Tuning, Prompt Tuning and Instruction Tuning.
In this context, the proposed method represents a fine-tuning task according to this categorization, as described above.
Other examples of fine-tuning tasks include Chen et al.~\cite{chen2020big}, who use large networks for unsupervised pretraining followed by supervised fine-tuning.
Zhou et al.~\cite{zhou2017fine} introduces a method that combines active learning and transfer learning to fine-tune CNNs for biomedical imaging. 
Radford et al.~\cite{radford2018improving} first introduced fine-tuning ideas into the design of large language models. 
For further literature about Fine-Tuning, we refer to~\cite{aghajanyan2020fine, kumar2022fine, ruiz2023fine}.

\section{Methodology}
In this section, we present the technical details of the proposed approach to CT image reconstruction. 
We outline the mathematical framework underlying our methodology and describe the key components, including the deep neural network architecture, 
the process of fine-tuning and the adaption of the network.

\subsection{Problem Formulation}
CT imaging is the task of reconstructing a high-quality CT image from raw projection data obtained through X-ray measurements from various angles. 
The forward model of this process describes the relationship between the acquired projection data and the desired image. 
Let $y$ denote the raw projection data acquired by the imaging system, $x \in \mathbb{R}^{n\times n}$ represents the true CT image, 
and $R$ denotes the (discrete) Radon transform that models the X-ray attenuation process in parallel-beam geometry~\cite{kisner2012model}. 
Then, the forward model can be expressed in a simplified form as an additive model:
\begin{align*}
    y &= R(x) + \epsilon,
\end{align*}
where $\epsilon$ is the measurement error including electronic noise, quantum noise, scatter, 
and patient motion artifacts~\cite{sandborg1995ctprinciples}. 
For a detailed model description, we refer to~\cite{leuschner2021lodopab}.For the corresponding mathematical model using fan-beam geometry, we refer to~\cite{jiangsheng2004fanbeam}.
The goal of CT image reconstruction is to recover an estimate 
of $x$ given the projection data $y$ and knowledge of the system matrix $R$.
We focus in particular on the case of LDCT reconstruction, where the projection data $y$ is obtained using a low radiation dose. 
The challenge in this case is that, the lower the radiation dose the higher is the noise level in the projection data $y$, 
and because the inversion of $R$ is ill-posed, 
the noise is amplified in the reconstruction process.

\subsection{Overview of the Proposed Two-Stage Approach}
We employ a two-stage approach for LDCT image reconstruction. The first stage involves an initial reconstruction using the FBP method, 
which is an established analytical technique for CT image reconstruction, but known to amplify measurement noise.
The second stage, the LDCT image enhancement stage, consists of a neural network $G_\theta$ that maps 
the low-dose FBP-reconstructed image to an enhanced image.
Mathematically, the two stage approach can be represented as follows:
\begin{align}\label{eq:two_stage}
    \hat x &= G_{\theta} \circ \mathrm{FBP}(y).
\end{align}
where $G_\theta$ is a neural network with trainable parameters $\theta,$  $\mathrm{FBP}$ denotes the FBP and $\hat x$ the final reconstruction result.

In our approach, $G_\theta$ is a neural network designed for removing Gaussian noise from natural grayscale images, in particular, they are trained on non-CT-type data.
We emphasize that the Gaussian denoisers have been pretrained
in prior works and thus do not require additional training effort in the context of our approach.

Subsequently, we fine-tune these denoisers for the specific downstream task of enhancing CT images, 
utilizing pairs of LDCT images and their corresponding NDCT counterparts. 
Figure~\ref{fig:method} provides an overview,
with the details of the method discussed in the following subsections.

\subsection{Pretrained Gaussian Denoisers for Natural Images}
\label{sec:pretrained_gaussian_denoiers}
For the neural network $G_{\theta}$ in~\eqref{eq:two_stage},
we use one of the following three Gaussian denoisers which have all been pretrained on natural grayscale images
in their respective works:

\subsubsection{DRUNet}
DRUNet~\cite{zhang2021pnp} was originally designed for a Plug-and-Play framework. 
It integrates the efficiency of UNets for image-to-image translation 
with ResNets' augmented modeling capacity via stacked residual blocks.
The details reveal a design with four scales, each featuring identity skip connections, 
$2\times 2$ strided convolution downscaling, and $2\times 2$ transposed convolution upscaling operations.
DRUNet is a deep neural network with 64, 128, 256, and 512 channels from its first to its fourth scale.
The initial and final convolutional layers, along with the strided and transposed convolutional layers, 
do not employ any activation function, which is inspired 
by the network architecture design for super-resolution outlined in~\cite{lim2017enhanced}.
The DRUNet architecture comprises 32.638.656 trainable parameters.
For Gaussian grayscale image denoising, the model is trained on a large-scale, 
diverse dataset comprising 8794 natural images 
derived from Berkeley segmentation, Waterloo Exploration, DIV2K, and Flick2K datasets. 
The optimization process involves minimizing the mean absolute error (MAE) loss between 
the denoised image and the ground truth, using the ADAM algorithm~\cite{kingma2017adam}.

\subsubsection{KBNet}
KBNet~\cite{zhang2023kbnet} utilizes a kernel basis attention module 
to adaptively aggregate spatial information through learnable kernel bases, 
capturing various image patterns that are fused by pixel-wise coefficients. 
This approach combines the strengths of CNNs and transformers, 
ensuring efficient and adaptive spatial context processing. 
Additionally, the multi-axis feature fusion block in KBNet 
extracts and integrates diverse features for enhanced image restoration. 
KBNet has 141.964.353 trainable parameters and is trained on the ImageNet validation dataset, comprising 50000 images, 
across three noise levels ($\sigma \in \{15, 25, 50\}$) for Gaussian grayscale image denoising.
The optimization process involves minimizing a composite loss function that combines the MAE loss and Structural Similarity (SSIM) loss, using the ADAM algorithm. 

\subsubsection{Restormer}
Restormer~\cite{zamir2022restormer} is designed to effectively restore high-resolution images 
using a Transformer architecture. 
Instead of traditional self-attention mechanisms, 
it introduces innovative techniques to manage computational complexity 
and enhance feature representation. 
Specifically, it utilizes a modified multi-head self-attention mechanism
that operates efficiently across channels rather than spatial dimensions, 
thereby capturing global context while reducing computational overhead. 
Additionally, Restormer incorporates a gated feed-forward network
that integrates gating mechanisms and depth-wise convolutions 
to enhance local image details and overall feature representation.
Restormer has 26.109.076 trainable parameters and is trained for Gaussian grayscale image denoising 
across three noise levels ($\sigma \in {15, 25, 50}$) 
using a combined dataset of 8594 images drawn 
from DIV2K, Flickr2K, BSD500, and the Waterloo Exploration dataset.

\subsection{Adaptation for the Downstream Task of LDCT Image Enhancement}
\label{sec:adaption}

\subsubsection{DRUNet}
\begin{figure}
    \centering
    \begin{subfigure}{0.4\textwidth}
        \centering
        \includegraphics[width=\linewidth]{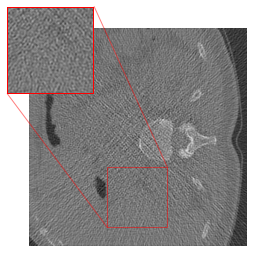}
        \caption{Low-dose FBP}
        \label{fig:input}
    \end{subfigure}
    \begin{subfigure}{0.4\textwidth}
        \centering
        \includegraphics[width=\linewidth]{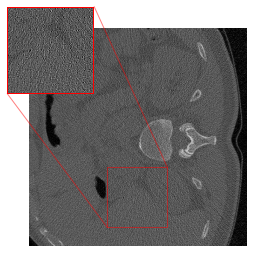}
        \caption{Ground Truth}
        \label{fig:gt}
    \end{subfigure}
    
    \medskip
    
    \begin{subfigure}{0.4\textwidth}
        \centering
        \includegraphics[width=\linewidth]{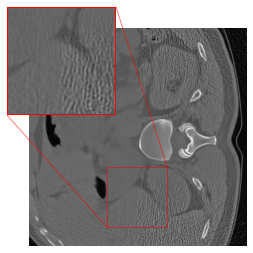}
        \caption{Output (Cropping)}
        \label{fig:output_cropping}
    \end{subfigure}
    \begin{subfigure}{0.4\textwidth}
        \centering
        \includegraphics[width=\linewidth]{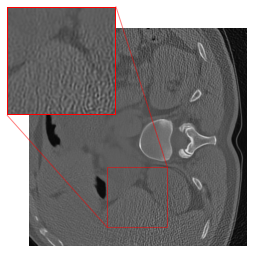}
        \caption{Output (Padding)}
        \label{fig:output_padding}
    \end{subfigure}
    \caption{An example highlighting the distinction between the DRUNet and FFDNet methodology of cropping versus the adapted padding strategy. (a) shows the low-dose FBP, while (b) displays the corresponding ground truth. (c) highlights the artifacts at the patch boundaries resulting from denoising on sub-images, whereas (d) exhibits artifact mitigation achieved through ommiting the partitioning into blocks.}
    \label{fig:cropping_padding_differences}
\end{figure}

The methodology from DRUNet and its predecessor FFDNet~\cite{zhang2018ffdnet} involves segmenting the input image into four distinct corner parts, 
each processed separately through the network. 
Zhang et al.~\cite{zhang2021pnp} utilize images with identically distributed Gaussian noise, allowing for individual denoising of sub-images and recombination without noticeable artifacts at the patch boundaries.
However, our application of this strategy in CT results in artifacts at the patch boundaries (cf. Figure~\ref{fig:output_cropping}), showcasing a sudden shift from intense to subtle smoothing with the cropping strategy. 
This issue may arise because the errors in CT images are not simply Gaussian noise, 
but rather a more complex mixture of noise types, 
including Poisson noise that has been log-transformed 
and then backpropagated via FBP.
For mitigating the artifacts, we modify the cropping approach of~\cite{zhang2018ffdnet}
by implementing padding in the output of the FBP to align with the network architecture. 
The DRUNet architecture operates solely on images sized as multiples of $2^4=16$ due to its four downsampling steps, whereas the LoDoPaB-CT dataset comprises images of size $362\times 362$, which does not conform to this requirement. 
To address this, we apply mirror padding to enlarge the images to $368\times 368$ before passing them through the network and subsequently cropping them back to the original size of $362\times 362.$ 
Figure~\ref{fig:cropping_padding_differences} illustrates the suitability of the proposed strategy.
(We point out that the image depicted in Figure~\ref{fig:cropping_padding_differences} represents one of the most challenging reconstructions within the dataset.)
For the 2016LDCTGC dataset, the images are sized at $512 \times 512$, 
which means they do not need additional padding and can be processed directly by the network.

Furthermore, the DRUNet includes an additional noise level map that is processed by the neural network. 
The noise level parameter within this map remains constant across the dataset during the training, validation, and testing phases.
This map is uniformly filled with a noise standard deviation parameter $\sigma$ which was initially used in~\cite{zhang2021pnp}
to adapt to different noise levels of the input image.
In this work, we do not use this adaption, and set the value to 1.

\subsubsection{KBNet and Restormer}

The Restormer model~\cite{zamir2022restormer} requires image dimensions to be multiples of 8. Consequently, a similar approach to that used with DRUNet was employed to enlarge the LoDoPaB-CT images to $368 \times 368$ pixels by applying mirror padding. 
While the KBNet model~\cite{zhang2023kbnet} has no such size constraint, we also resized its LoDoPaB-CT input images to $368 \times 368$ 
for consistency and better comparability.
For the 2016LDCTGC dataset, as in the case of DRUNet,
both networks can process the images directly.

\subsubsection{Loss Function}
\label{sec:loss_function}
Following the underlying neural network methodology of Zhang et al.~in~\cite{zhang2021pnp}, we initially experimented with the MAE loss function.
The MAE loss emphasizes pixel-wise intensity accuracy and provides a basic approach to improve CT image reconstruction. 
MAE is generally considered more robust to outliers than MSE~\cite{bishop2006pattern}.
The MAE loss, denoted as $L_{\mathrm{MAE}}$, is given as follows:
\begin{center}
    $L_{\mathrm{MAE}}=\frac{1}{N}\sum_{i=1}^{N} |\hat{x}_i - y_i|$
\end{center}
Here, $N$ represents the total number of pixels in an image, $\hat{x}_i$ denotes the $i$-th pixel of the output obtained from the neural network
while $y_i$ denotes the $i$-th pixel of the ground truth NDCT image.

Inspired by~\cite{xiong2023reunet} and~\cite{zhang2023kbnet} we use a composite loss function combining the MAE loss and the Structural Similarity (SSIM) loss~\cite{wang2004ssim}. 
The SSIM measures the structural similarity between two images by taking into account statistical properties of local neighborhoods instead of pixel-wise intensity accuracy. 
According to~\cite{setiadi2021psnrvsssim}, SSIM is more adept at capturing perceptually induced changes that are frequently overlooked by traditional metrics. 
This is attributed to its incorporation of perceptual psychological phenomena, including brightness and contrast masking terms.
To promote stability and generalization, we also incorporate $L_2$ regularization into the training process.
The resulting regularized SSIM-based loss function $L$, was employed for training all three networks. Its expression is given by the following formula:
\begin{align}
    L&=L_{\mathrm{MAE}}+\alpha (1 - \SSIM) + \lambda \|\theta\|_2^2
\end{align}
The calculation of $\SSIM$ follows the methodology outlined in~\cite{wang2004ssim}, adressing both structural and perceptual similarity between images. 
The parameter $\alpha > 0$ determines the weight of the $(1-\SSIM)$ term.
Additionally, $\|\theta\|_2^2$ represents the squared $L_2$ norm of the model parameters
and $\lambda > 0 $ denotes the regularization hyperparameter.

\section{Experiments}

\subsection{Experimental Setup}
\label{exp_setup}

\subsubsection{Implementation Details}
\label{imp_det}

For training our model, we employed the ADAM optimizer \cite{kingma2017adam}. The hyperparameters $\beta_1$ and $\beta_2$ in the ADAM algorithm were set to their default values of 0.9 and 0.999, respectively, balancing the gradient and squared gradient moving averages.
If not stated differently, we set the parameters $\alpha$ to $5$ and $\lambda$ to $10^{-5}$.
Additionally, for the learning rate strategy, an initial value of $10^{-4}$ was chosen, 
and it was halved after a certain amount of epochs during our training.
Our implementation was realized using Python 3.9.16, and the PyTorch 2.0.0 library served as the foundational framework for our algorithm. In both the training and testing phases, we utilized an NVIDIA A40 GPU equipped with 40 GB RAM.

We employ two types of data augmentation to enhance the robustness of our network. 
The first method involves rotating the input and ground truth images in all four major orientations ($0^\circ, 90^\circ, 180^\circ, 270^\circ$). 
This procedure aims to enhance the network’s robustness to these transformations 
by providing it with diverse orientations of the data. 
The second augmentation technique involves the addition of Gaussian noise to the training images. 
More precisely, the entire dataset is utilized in two ways: initially in its original state, and subsequently, with the addition of 1\% Gaussian noise to the input images. 
By augmenting the training data with synthetic noise, we expose the network to a wider spectrum of noise patterns, enabling it to learn to distinguish between inherent anatomical structures and noise-induced artifacts.

\subsubsection{The LoDoPaB-CT Dataset}

The LoDoPaB-CT dataset~\cite{leuschner2021lodopabcomp} comprises 46573 pairs
of human chest CT images and their corresponding simulated low-dose measurements.
These images are sourced from a spectrum of imaging scenarios, 
providing a robust benchmark for assessing the effectiveness of our approach under varying conditions. 
The dataset's ground truth images are derived from the LIDC/IDRI dataset~\cite{armato2011lidc}, 
specifically cropped to dimensions of $362\times362$ pixels.
This dataset is split into distinct sets: 35820 training samples, 3522 validation samples, 3553 test samples, and 3678 samples for the
challenge phase. The ground truth images of the challenge dataset remain confidential and 
are exclusively used for evaluation via the LoDoPaB-CT challenge's online submission system
(cf.~\url{https://lodopab.grand-challenge.org}).
In~\cite{leuschner2021lodopab} various deep learning-based image reconstruction algorithms 
were systematically evaluated on the LoDoPaB-CT dataset. 
This renders the dataset an established and recognized benchmark for quantitatively 
comparing different approaches in the context of LDCT reconstruction.

\subsubsection{2016 Low Dose CT Grand Challenge (2016LDCTGC) Dataset}
The 2016LDCTGC dataset~\cite{mccollough2017low} 
is a comprehensive collection of medical imaging data designed to advance LDCT imaging techniques. 
This dataset was released as part of the Low Dose CT Grand Challenge, 
an initiative organized by the Mayo Clinic 
and the American Association of Physicists in Medicine (AAPM). 
The CT projection data, acquired using a helical scan procedure with a fan-beam geometry, is available for both full and simulated lower dose levels. The corresponding CT images for full-dose projections are reconstructed using commercial CT systems. 
For scans conducted on the SOMATOM Definition Flash CT scanner from Siemens Healthcare, 
reconstructed images of dimensions $512\times512$ from lower dose projections are also provided. 
The dataset comprises a total of 299 scans, including 99 head, 100 chest, and 100 abdomen scans.
To ensure a homogeneous sample, we exclusively use chest scans acquired with the SOMATOM Definition Flash CT scanner for our experiments. Since low-dose reconstructions are already provided for this data, the first stage of the proposed approach (FBP) is unnecessary, as the data is already in the image domain.

\subsubsection{Evaluation Metrics}
We utilize the metrics proposed in the LoDoPaB-CT challenge \cite{leuschner2021lodopabcomp},
which are the following four metrics: peak signal-to-noise ratio (PSNR) and structural similarity (SSIM),
as well as their corresponding fixed range (FR) variants, PSNR-FR and SSIM-FR, which adopt a fixed range rather 
than the difference between the highest and lowest values in the ground truth image. 
PSNR quantifies the ratio between the maximum signal intensity and noise affecting visualization \cite{sara2019psnr}. 
The SSIM introduced by \cite{wang2004ssim} measures the structural similarity between two images, as described in Section \ref{sec:loss_function}.
It is important to note that both PSNR and SSIM have limitations in capturing certain types of artifacts that may impact clinical interpretation. Nevertheless, they remain valuable tools for providing quantitative assessments of image quality.
In all metrics, a higher value indicates a better reconstruction quality.

\subsection{Experimental Results}
\label{exp_results}

\subsubsection{Efficiency of the proposed approach}
\label{lodopab_challenge_results}

\begin{table*}
    \centering
    \tiny
    \begin{tabular*}{\textwidth}{@{\extracolsep{\fill}}p{2cm}cccccc}
        \toprule
        & \textbf{Mean Pos.} & \textbf{PSNR} & \textbf{PSNR-FR} & \textbf{SSIM} & \textbf{SSIM-FR} \\
        \midrule
        \textbf{FBP} & $54$ & $30.19 \pm 2.555$	& $34.46 \pm 2.182$ & $0.7274 \pm 0.1273$ & $0.8356 \pm 0.08497$ \\
        \textbf{TV} & $48.5$ & $33.36 \pm 2.736$	& $37.63 \pm 2.702$ & $0.8303 \pm 0.1211$ & $0.9030 \pm 0.08196$ \\
        \textbf{CINN~\cite{denker2020ldct}} & $34.8$ & $35.54 \pm 3.509$ & $39.81 \pm 3.477$ & $0.8542 \pm 0.1215$ & $0.9189 \pm 0.08143$ \\
        \textbf{UNet++~\cite{zhou2018unetpp}} & $31$ & $35.37 \pm 3.361$ & $39.64 \pm 3.404$ & $0.8609 \pm 0.1192$ & $0.9225 \pm 0.08007$ \\
        \textbf{MS-D-CNN~\cite{pelt2018msdcnn}} & $30.3$ & $35.85 \pm 3.601$ & $40.12 \pm 3.557$ & $0.8576 \pm 0.1221$ & $0.9210 \pm 0.08161$ \\
        \textbf{Original~UNet~\cite{ronneberger2015unet}} & $28$ & $35.87 \pm 3.593$ & $40.14 \pm 3.566$ & $0.8592 \pm 0.1200$ & $0.9218 \pm 0.08021$ \\
        \textbf{LoDoPaB~UNet~\cite{leuschner2021lodopab}} & $20.3$ & $36.00 \pm 3.631$ & $40.28 \pm 3.590$ & $0.8618 \pm 0.1185$ & $0.9233 \pm 0.07912$ \\
        \textbf{MUSC~\cite{liu2022musc}} & $16.5$ & $36.08 \pm 3.682$ & $40.35 \pm 3.644$ & $0.8628 \pm 0.1193$ & $0.9238 \pm 0.07957$ \\
        \textbf{ISTA~UNet~\cite{liu2020istaunet}} & $16.3$ & $36.09 \pm 3.687$ & $40.36 \pm 3.648$ & $0.8623 \pm 0.1197$ & $0.9236 \pm 0.07969$ \\
        \textbf{\mbox{Learned~P.-D.~\cite{adler2018learnedprimaldual}}} & $11.5$ & $36.25 \pm 3.696$ & $40.52 \pm 3.635$ & $0.8662 \pm 0.1152$ & $0.9260 \pm 0.07635$ \\
        \textbf{\mbox{FBP-DTSGD~FS}} & $3.3$ & $36.41 \pm 3.896$ & $40.68 \pm 3.871$ & $\mathit{0.8699 \pm 0.1160}$ & $\mathit{0.9274 \pm 0.07819}$ \\
        \textbf{ItNet~\cite{genzel2022nearexactrecovery}} & $2$ & $\mathbf{36.47 \pm 3.791}$ & $\mathbf{40.74 \pm 3.743}$ & $0.8692 \pm 0.1150$ & $\mathbf{0.9275 \pm 0.07632}$ \\
        \textbf{\mbox{FBP-DTSGD~PT}} & $1.8$ & $\mathit{36.43 \pm 3.894}$ & $\mathit{40.70 \pm 3.867}$ & $\mathbf{0.8700 \pm 0.1162}$ & $\mathit{0.9274 \pm 0.07819}$ \\
        \bottomrule
    \end{tabular*}
    \caption{Evaluation of our method on the LoDoPaB-CT challenge dataset, compared to various state-of-the-art methods. 
    The metrics are taken from~\cite{leuschner2021lodopabcomp}. Mean Position, in this context, refers to the average placement across the four distinct positions on the leaderboard concerning the four evaluation metrics.
    The results are taken from the challenge platform (\url{https://lodopab.grand-challenge.org/evaluation/challenge/leaderboard}) on March 4th, 2025. The highest overall score is emphasized in boldface, while the second-best score is highlighted in italic.
    }
\label{tab:method_comparison}
\end{table*}

We evaluated the DRUNet variant of our proposed two-stage method FBP-DTSGD
within the LoDoPaB-CT challenge.
The training was performed
as described in Section~\ref{imp_det}, 
including rotational augmentation and Gaussian noise augmentation. 
In this training, we utilize the pretrained DRUNet weights, which were optimized for Gaussian noise removal on natural grayscale images in a prior work.
Initially, we trained the DRUNet exclusively on the LoDoPaB-CT training dataset for 264 epochs. Following this, we extended the training to include the entire publicly available LoDoPaB-CT dataset (training, testing, and validation subsets) for an additional 264 epochs, withholding only a subset of 271 images for testing purposes.
The final model was then submitted to the challenge platform for evaluation 
on the not publicly available challenge set.
At the time of writing (on March 4th, 2025), 
as documented on the challenge platform (\url{https://lodopab.grand-challenge.org/evaluation/challenge/leaderboard}), our approach now holds the overall first mean position, surpassing the unrolled iterative methods ItNet and Learned Primal-Dual. While our method achieves the highest rank in the SSIM metric, ItNet has a slight advantage in the PSNR, PSNR-FR, and SSIM-FR metrics.
For a detailed comparison of our method and others, 
including references to corresponding papers, 
we provide an overview in Table~\ref{tab:method_comparison}. 
In this table, our method is denoted as \enquote{FBP-DTSGD PT}, where \enquote{PT} stands for \enquote{pretrained}.

Table~\ref{tab:inference_time} presents the inference times (in seconds) for the FBP method and our proposed FBP-DTSGD approach, using the three Gaussian denoisers: DRUNet, KBNet, and Restormer.
The times refer to processing a single image of the LoDoPaB-CT dataset with a NVIDIA A40 GPU with 40 GB of RAM.
The inference time of the DRUNet model is only a fraction of the FBP reconstruction time, which makes the FBP the domiant part of the processing time for the complete FBP-DTSGD method using the DRUNet model. 
For the KBNet and Restormer, the network inference time is at most two times the FBP reconstruction time.
Furthermore, the inference time of the second place of the LoDoPaB-CT challenge, ItNet, is evaluated. 
ItNet executes a block consisting of a UNet, a forward operator and a FBP five times. Moreover, the final reconstruction of ItNet is computed by an ensemble of ten networks due to statistical fluctuations~\cite{genzel2022nearexactrecovery}. Consequently, ItNet takes over 60 times longer to reconstruct a single image compared to the DRUNet variant of our method, FBP-DTSGD, using this particular hardware (cf.~Table~\ref{tab:inference_time}).

\begin{table}
    \centering
    \begin{tabular}{lrrr}
    \toprule
    \textbf{Method} & \textbf{Runtime (s)} \\
    \midrule
    FBP       & 0.09375 \\
    ItNet     & 6.98446 \\
    FBP-DTSGD (DRUNet)    & 0.10938 \\
    FBP-DTSGD (KBNet)     & 0.21876 \\
    FBP-DTSGD (Restormer) & 0.28125 \\
    
    \bottomrule
    \end{tabular}
    \caption{Runtime (in seconds) for the FBP method and inference times (in seconds) for the three Gaussian denoisers (DRUNet, KBNet, Restormer), which form the two stages of our proposed method, FBP-DTSGD. For comparison, the inference time of the state-of-the-art method ItNet is also included. All times correspond to processing a single image from the LoDoPaB-CT dataset using an NVIDIA A40 GPU with 40 GB of RAM.}
    \label{tab:inference_time}
\end{table}

\subsubsection{Ablation Study}

\paragraph{\it{Effectiveness of a Gaussian Denoiser for LDCT Image Enhancement}}

We investigate how the architecture of Gaussian denoiser models affect the performance of the proposed FBP-DTSGD method. To this end, we explore different architectures, including transformer-based models. In particular, we evaluate three Gaussian denoiser models -- DRUNet, KBNet, and Restormer (cf.~\ref{sec:pretrained_gaussian_denoiers}) -- within the FBP-DTSGD framework.
We evaluate the models on small subsets of the LoDoPaB-CT and 2016LDCTGC datasets. For the LoDoPaB-CT dataset, we used 0.2\%, 1\%, and 5\% of the training and validation sets, corresponding to 71, 358, and 1791 training images, respectively. Testing was performed on 300 randomly selected images from the LoDoPaB-CT test dataset. This setup allows us to conduct a series of experiments with the given computational power and to study the effects of limited training data. For the 2016LDCTGC dataset, we selected the same number of images for training, validation, and testing, all chosen randomly. The networks were trained for 96 epochs.

The first three rows of Table~\ref{tab:denoiser_comparison_lodopab} summarize the performance of these three neural network architectures on subsets of the LoDoPaB-CT dataset. The results for the 2016LDCTGC dataset are presented in the first three rows of Table~\ref{tab:denoiser_comparison_2016ldctgc}. The impact of pretrained networks versus networks trained from scratch will be discussed in the following paragraph.

\begin{table*}
    \centering
    \begin{tabular}{p{5cm}cc}
        \toprule
        \textbf{LoDoPaB-CT Dataset} & \textbf{PSNR} & \textbf{SSIM} \\
        \midrule
        \multicolumn{3}{c}{\textbf{71 Training Images}} \\
        \midrule
        \textbf{DRUNet Pretrained} & $35.30 \pm 4.344$ & $0.8374 \pm 0.1506$ \\
        \textbf{KBNet Pretrained} & $35.56 \pm 4.515$ & $0.8419 \pm 0.1510$ \\
        \textbf{Restormer Pretrained} & $35.62 \pm 4.531$ & $0.8436 \pm 0.1494$ \\
        \textbf{DRUNet From Scratch} & $34.39 \pm 4.097$ & $0.8185 \pm 0.1508$ \\
        \textbf{KBNet From Scratch} & $34.26 \pm 4.011$ & $0.8185 \pm 0.1533$ \\
        \textbf{Restormer From Scratch} & $34.84 \pm 4.191$ & $0.8319 \pm 0.1533$ \\
        \midrule
        \multicolumn{3}{c}{\textbf{358 Training Images}} \\
        \midrule
        \textbf{DRUNet Pretrained} & $35.87 \pm 4.643$ & $0.8468 \pm 0.1499$ \\
        \textbf{KBNet Pretrained} & $35.93 \pm 4.662$ & $0.8482 \pm 0.1494$ \\
        \textbf{Restormer Pretrained} & $36.02 \pm 4.717$ & $0.8510 \pm 0.1471$ \\
        \textbf{DRUNet From Scratch} & $35.35 \pm 4.432$ & $0.8416 \pm 0.1505$ \\
        \textbf{KBNet From Scratch} & $35.23 \pm 4.367$ & $0.8392 \pm 0.1523$ \\
        \textbf{Restormer From Scratch} & $35.70 \pm 4.535$ & $0.8465 \pm 0.1503$ \\
        \midrule
        \multicolumn{3}{c}{\textbf{1791 Training Images}} \\
        \midrule
        \textbf{DRUNet Pretrained} & $36.14 \pm 4.782$ & $0.8516 \pm 0.1482$ \\
        \textbf{KBNet Pretrained} & $36.03 \pm 4.702$ & $0.8512 \pm 0.1474$ \\
        \textbf{Restormer Pretrained} & $36.16 \pm 4.794$ & $0.8540 \pm 0.1450$ \\
        \textbf{DRUNet From Scratch} & $35.90 \pm 4.696$ & $0.8499 \pm 0.1485$ \\
        \textbf{KBNet From Scratch} & $35.84 \pm 4.628$ & $0.8501 \pm 0.1476$ \\
        \textbf{Restormer From Scratch} & $36.07 \pm 4.747$ & $0.8520 \pm 0.1472$ \\
        \bottomrule
    \end{tabular}
    \caption{Performance comparison based on PSNR and SSIM metrics for DRUNet, KBNet, and Restormer, trained from scratch and pretrained, across three different subsets of the LoDoPaB-CT dataset.}
    \label{tab:denoiser_comparison_lodopab}
\end{table*}

\begin{table*}
    \centering
    \begin{tabular*}{\textwidth}{@{\extracolsep{\fill}}p{5cm}cc}
        \toprule
        \textbf{2016LDCTGC Dataset} & \textbf{PSNR} & \textbf{SSIM} \\
        \midrule
        \multicolumn{3}{c}{\textbf{71 Training Images}} \\
        \midrule
        \textbf{DRUNet Pretrained} & $30.96 \pm 2.694$ & $0.7703 \pm 0.08088$ \\
        \textbf{KBNet Pretrained} & $31.34 \pm 3.087$ & $0.7752 \pm 0.08207$ \\
        \textbf{Restormer Pretrained} & $31.42 \pm 3.079$ & $0.7760 \pm 0.08170$ \\
        \textbf{DRUNet From Scratch} & $29.53 \pm 2.417$ & $0.7478 \pm 0.08405$ \\
        \textbf{KBNet From Scratch} & $30.32 \pm 2.860$ & $0.7577 \pm 0.08304$ \\
        \textbf{Restormer From Scratch} & $30.89 \pm 2.839$ & $0.7640 \pm 0.08292$ \\
        \midrule
        \multicolumn{3}{c}{\textbf{358 Training Images}} \\
        \midrule
        \textbf{DRUNet Pretrained} & $31.50 \pm 2.721$ & $0.7780 \pm 0.08063$ \\
        \textbf{KBNet Pretrained} & $32.07 \pm 2.703$ & $0.7801 \pm 0.08109$ \\
        \textbf{Restormer Pretrained} & $32.65 \pm 2.493$ & $0.7827 \pm 0.08109$ \\
        \textbf{DRUNet From Scratch} & $30.46 \pm 2.708$ & $0.7623 \pm 0.08263$ \\
        \textbf{KBNet From Scratch} & $31.22 \pm 2.736$ & $0.7722 \pm 0.08152$ \\
        \textbf{Restormer From Scratch} & $31.92 \pm 2.771$ & $0.7794 \pm 0.08119$ \\
        \midrule
        \multicolumn{3}{c}{\textbf{1791 Training Images}} \\
        \midrule
        \textbf{DRUNet Pretrained} & $32.22 \pm 2.603$ & $0.7837 \pm 0.08087$ \\
        \textbf{KBNet Pretrained} & $32.98 \pm 2.462$ & $0.7842 \pm 0.08109$ \\
        \textbf{Restormer Pretrained} & $33.06 \pm 2.445$ & $0.7864 \pm 0.08102$ \\
        \textbf{DRUNet From Scratch} & $31.31 \pm 2.578$ & $0.7805 \pm 0.08068$ \\
        \textbf{KBNet From Scratch} & $32.56 \pm 2.505$ & $0.7817 \pm 0.08103$ \\
        \textbf{Restormer From Scratch} & $33.04 \pm 2.504$ & $0.7854 \pm 0.08117$ \\
        \bottomrule
    \end{tabular*}
    \caption{Performance comparison based on PSNR and SSIM metrics for DRUNet, KBNet, and Restormer, trained from scratch and pretrained, across three different subsets of the 2016LDCTGC dataset.}
    \label{tab:denoiser_comparison_2016ldctgc}
\end{table*}

\begin{figure*}
    \centering
    
    % First row
    \begin{minipage}{0.19\textwidth}
        \centering
        \textbf{Ground Truth} \\[0.5ex]
        \includegraphics[width=\textwidth]{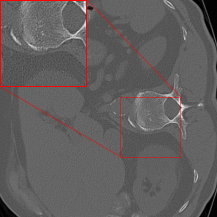} \\
        \small{$\mathrm{PSNR} = \infty$\\$\SSIM = 1.0$} % Replace X and Y with the actual values
    \end{minipage} \hfill
    \begin{minipage}{0.19\textwidth}
        \centering
        \textbf{FBP}\\[3.3ex]
        \includegraphics[width=\textwidth]{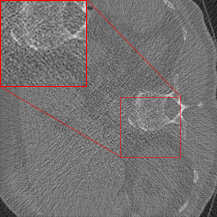} \\
        \small{$\mathrm{PSNR} = 21.47$\\$\SSIM = 0.2372$}
    \end{minipage} \hfill
    \begin{minipage}{0.19\textwidth}
        \centering
        \textbf{Output DRUNet} \\[0.5ex]
        \includegraphics[width=\textwidth]{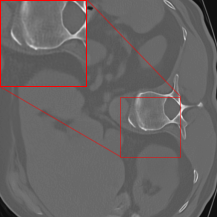} \\
        \small{$\mathrm{PSNR} = 30.41$\\$\SSIM = 0.6375$}
    \end{minipage} \hfill
    \begin{minipage}{0.19\textwidth}
        \centering
        \textbf{Output KBNet} \\[0.5ex]
        \includegraphics[width=\textwidth]{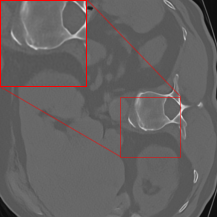} \\
        \small{$\mathrm{PSNR} = 30.30$\\$\SSIM = 0.6332$}
    \end{minipage} \hfill
    \begin{minipage}{0.19\textwidth}
        \centering
        \textbf{Output Restormer} \\[0.5ex]
        \includegraphics[width=\textwidth]{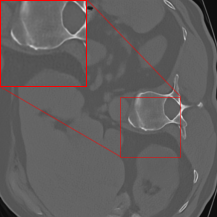} \\
        \small{$\mathrm{PSNR} = 30.31$\\$\SSIM = 0.6361$}
    \end{minipage} \\[1ex]
    
    % Second row
    \begin{minipage}{0.19\textwidth}
        \centering
        \includegraphics[width=\textwidth]{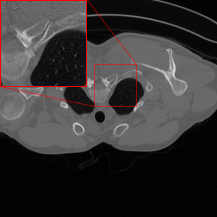} \\
        \small{$\mathrm{PSNR} = \infty$\\$\SSIM = 1.0$}
    \end{minipage} \hfill
    \begin{minipage}{0.19\textwidth}
        \centering
        \includegraphics[width=\textwidth]{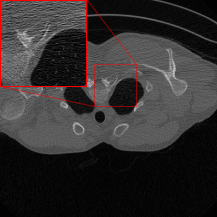} \\
        \small{$\mathrm{PSNR} = 23.39$\\$\SSIM = 0.4382$}
    \end{minipage} \hfill
    \begin{minipage}{0.19\textwidth}
        \centering
        \includegraphics[width=\textwidth]{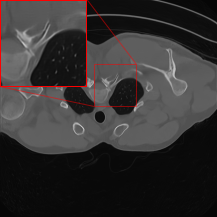} \\
        \small{$\mathrm{PSNR} = 26.98$\\$\SSIM = 0.7782$}
    \end{minipage} \hfill
    \begin{minipage}{0.19\textwidth}
        \centering
        \includegraphics[width=\textwidth]{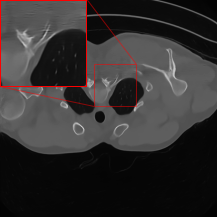} \\
        \small{$\mathrm{PSNR} = 26.74$\\$\SSIM = 0.7662$}
    \end{minipage} \hfill
    \begin{minipage}{0.19\textwidth}
        \centering
        \includegraphics[width=\textwidth]{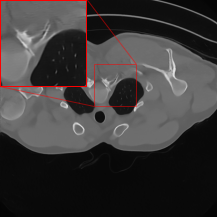} \\
        \small{$\mathrm{PSNR} = 26.92$\\$\SSIM = 0.7681$}
    \end{minipage}
    
    \caption{Reconstruction results for two test samples from the LoDoPaB-CT test dataset (top row) and the 2016LDCTGC dataset (bottom row). The first column displays the ground truth, followed by the FBP in the second column. The next three columns present reconstructions by the Gaussian denoisers DRUNet, KBNet, and Restormer. Each pretrained denoiser model was trained for 96 epochs on 1791 image pairs from either the LoDoPaB-CT training dataset or the 2016LDCTGC dataset. The proposed two-stage method effectively enhances the FBP reconstructions in each case, as demonstrated by the PSNR and SSIM metrics provided below each subimage.}
    \label{fig:denoiser_comparison}
\end{figure*}

Figure~\ref{fig:denoiser_comparison} shows reconstruction results from the three networks for a test sample from the LoDoPaB-CT test dataset and one from the 2016LDCTGC dataset. All trained networks successfully improve the FBP reconstructions, as shown by the improved image quality and finer structural details.

When comparing the training times of the three networks, 
the DRUNet was the most computationally efficient. 
When training on 71 images from the LoDoPaB-CT dataset 
over 96 epochs, the DRUNet took 1026 seconds, 
while the KBNet and Restormer required 5390 seconds 
and 5252 seconds, respectively.

\paragraph{\it{Effectiveness of Using Pretrained Weights}}

We explore the impact of using pretrained weights from a different domain and task -- Gaussian noise removal from natural grayscale images -- for LDCT image enhancement.
To this end, we compare the performance of our proposed two-stage method, FBP-DTSGD, against the same networks trained from scratch.

As previously discussed, we evaluated three different Gaussian denoisers (DRUNet, KBNet, Restormer) on two distinct datasets (LoDoPaB-CT dataset, 2016LDCTGC dataset).
The first three rows of Table~\ref{tab:denoiser_comparison_lodopab} and Table~\ref{tab:denoiser_comparison_2016ldctgc} present the results using pretrained weights, while the last three rows display the results using the same network architectures with randomly initialized weights.
Our findings indicate that fine-tuning the pretrained network 
consistently yields better metrics than training from scratch across all tested scenarios. 
Additionally, the advantages of pretraining become more pronounced 
as the number of images used for fine-tuning decreases. 
Notably, training the network from scratch with 358 images produces 
results comparable to fine-tuning the pretrained network with only 71 images. 
Similarly, training from scratch with 1791 images achieves 
comparable results to fine-tuning with 358 images. 
This demonstrates that less task-specific data is needed 
to achieve similar outcomes.

\begin{figure}
    \centering
    \begin{subfigure}{0.49\textwidth}
        \centering
        \includegraphics[width=\linewidth]{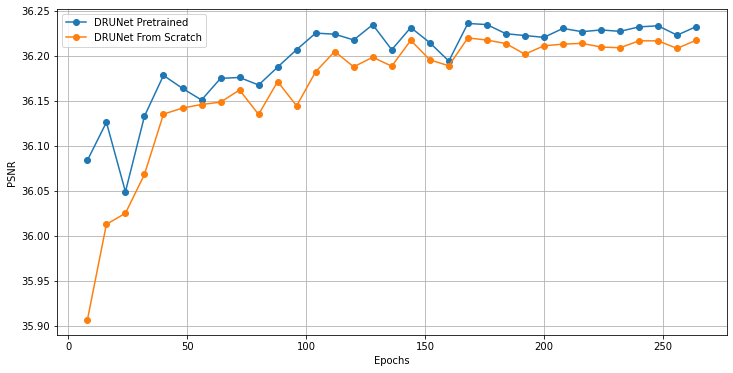}
        \caption{PSNR Comparison}
        \label{fig:psnr}
    \end{subfigure}
    \hfill
    \begin{subfigure}{0.49\textwidth}
        \centering
        \includegraphics[width=\linewidth]{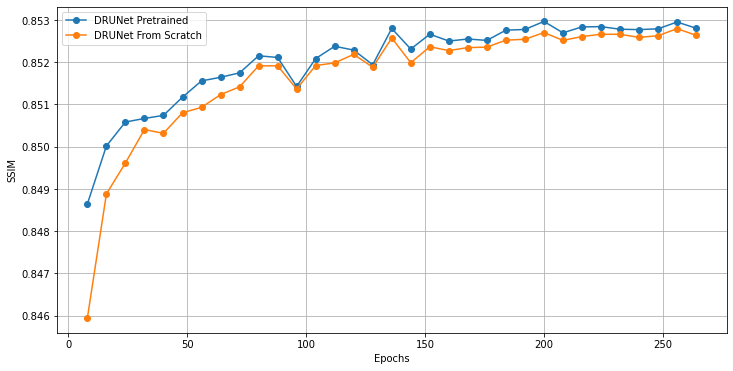}
        \caption{SSIM Comparison}
        \label{fig:ssim}
    \end{subfigure}
    
    \caption{Comparison of PSNR and SSIM metrics for the pretrained DRUNet model and the DRUNet model trained from scratch on the LoDoPaB-CT test dataset. (a) shows the comparison of PSNR values, and (b) displays the comparison of the SSIM values.}
    \label{fig:psnr_ssim_comparison}
\end{figure}

Next, we extended our experiments to the entire LoDoPaB-CT dataset. 
To maintain a manageable computational workload, 
we focused exclusively on the DRUNet variant of our method, 
given its lower computational cost compared to the other two Gaussian denoisers we evaluated.
We note that -- even though the LoDoPaB-CT training dataset is larger than the dataset used for pretraining of the DRUNet --
we stick to the term \enquote{fine-tuning} for this setup.
One might ask whether training on the LoDoPaB-CT dataset \enquote{overwrites} the prior training on natural images thereby annihilating the benefit of using a pretrained network. 
To address this question, a DRUNet from scratch and a pretrained one were trained on the LoDoPaB-CT training dataset for a large number of 264 epochs.
Figure~\ref{fig:psnr_ssim_comparison} presents the PSNR and SSIM values on the public LoDoPaB-CT test dataset over multiple training epochs. The results show that the proposed method, using a pretrained DRUNet, consistently produces better results than a DRUNet trained from scratch across all epochs.
The reason for the better performance of the pretrained network might be that -- even though the characteristics of natural images and CT images are different --
the network was able to capture features during the pretraining stage that are useful for the LDCT image enhancement task.
In summary, this results in a higher \enquote{effective} training time
for the proposed method.
We again emphasize that the actual training load
for both networks (from scratch and pretrained) is identical for the task in this work, 
because the DRUNet has been trained in a different prior work and is thus already amortized.

Furthermore, we evaluated the DRUNet trained from scratch on the LoDoPaB-CT challenge, using the same training procedure as described in Section~\ref{lodopab_challenge_results}. 
The results, labeled as \enquote{FBP-DTSGD FS} (with \enquote{FS} denoting \enquote{from scratch}), are summarized in Table~\ref{tab:method_comparison}.
We observe that even after extensive training, the pretrained DRUNet continues to achieve slightly better performance.

\section{Discussion and Conclusion}

In this paper, we proposed FBP-DTSGD, a two-stage approach for LDCT image reconstruction. The first stage employs classical FBP reconstruction, while the second stage enhances the reconstructed images using a fine-tuned deep learning model.
The novelty lies in the use of existing Gaussian denoisers, initially trained on natural grayscale images, and adapting them to LDCT image enhancement through fine-tuning. This adaptation addresses both a domain shift from natural grayscale images to CT images and a task shift from Gaussian noise removal to LDCT image enhancement.

We evaluated FBP-DTSGD in the LoDoPaB-CT challenge, where it secured the first mean position overall, surpassing the unrolled iterative methods. Our method achieved the highest SSIM score, while ItNet holds a slight edge in the PSNR, PSNR-FR, and SSIM-FR metrics. In our ablation study, we investigated how the architecture of Gaussian denoiser models influence the performance and evaluated the impact of pretraining the models on a different domain and task. We evaluated three distinct Gaussian denoiser models: DRUNet~\cite{zhang2021pnp}, KBNet~\cite{zhang2023kbnet}, and Restormer~\cite{zamir2022restormer}. These pretrained networks were fine-tuned on various subsets of the LoDoPaB-CT and 2016LDCTGC datasets, using different amounts of training data.
The results showed that FBP-DTSGD works reliably with different Gaussian denoiser architectures. Furthermore, fine-tuning the pretrained networks produces better results than training the same networks from scratch, as evidenced by a higher PSNR and SSIM metric. Notably, as the amount of training data decreased, the benefits of our approach became even more significant, demonstrating that less task-specific data is required to achieve comparable results. Moreover, even after extensive training on the entire LoDoPaB-CT dataset, the pretrained DRUNet showed a slight edge over the DRUNet trained from scratch. While this improvement is incremental, it proved crucial in securing first mean position overall in the LoDoPaB-CT challenge.

Compared to the unrolled iterative approaches, such as ItNet and Learned-Primal-Dual~\cite{adler2018learnedprimaldual}, which are the top-ranking methods in the LoDoPaB-CT challenge, our architecture is less complex in the sense that unrolled iterative methods require multiple applications of both the forward imaging operator and FBP.
In contrast, our method consists of only a single FBP computation followed by a forward pass through a network. 
As a result, our method achieves significantly lower inference times than ItNet.

While the benefit of pretraining is well-known and utilized by various methods, a key distinction of our approach is that the pretraining is conducted entirely on natural grayscale images, without using domain-specific CT data. This sets our method apart from state-of-the-art approaches like ItNet~\cite{genzel2022nearexactrecovery}, which rely on domain-specific pretraining.
Additionally, the pretraining does not increase the training load for the proposed method, as the pretraining has been performed in a different prior work.
A further advantage is the flexibility of our approach: if future advancements yield more effective Gaussian denoisers, they can be easily integrated into our framework, potentially enhancing its performance.

We eventually try to give some intuition why it can be beneficial to use a Gaussian denoiser for LDCT image enhancement. The combination of FBP and a Gaussian denoiser can be seen as the first step in a plug-and-play framework, where the denoiser is applied iteratively alongside a data fidelity term to refine the FBP reconstruction~\cite{zhang2021pnp}. Our fine-tuning approach can be interpreted as an approximation of the residual between the ground truth and this initial reconstruction. In contrast, ItNet adopts a different strategy, addressing the same gap by unrolling an iterative algorithm.

One topic of future research consists of the application of the proposed approach to other imaging modalities. Since the method relies on CT data and employs FBP for the (regularized) inversion of the imaging operator, adapting it to a different modality would require adjustments to the reconstruction method and a suitable dataset for fine-tuning.

% Appendices
\appendix

\section{Additional Experiments}
The networks in the subsequent experiments are trained using the LoDoPaB-CT training dataset and evaluated on the LoDoPaB-CT test dataset, unless specified otherwise.

\subsection{Effectiveness of the Rotational Augmentation}
\begin{figure}
    \centering
    % First row (2 images)
    \begin{subfigure}{0.22\textwidth}
        \centering
        \includegraphics[width=\linewidth]{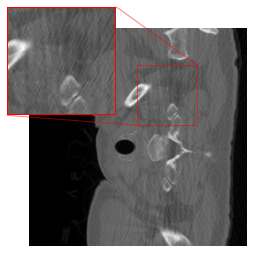}
        \caption{Ground Truth}
        \label{fig:rotaug_gt}
    \end{subfigure}
    \begin{subfigure}{0.22\textwidth}
        \centering
        \includegraphics[width=\linewidth]{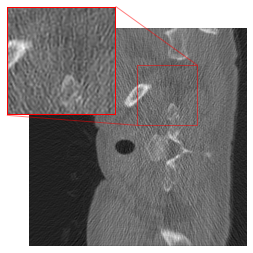}
        \caption{Low-Dose FBP}
        \label{fig:rotaug_input}
    \end{subfigure}
    
    \medskip
    
    % Second row (4 images)
    \begin{subfigure}{0.22\textwidth}
        \centering
        \includegraphics[width=\linewidth]{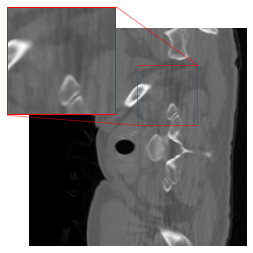}
        \caption{Output (0°) without Rotational Augmentation}
        \label{fig:rotaug_output_norot_0}
    \end{subfigure}
    \begin{subfigure}{0.22\textwidth}
        \centering
        \includegraphics[width=\linewidth]{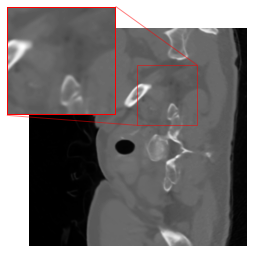}
        \caption{Output (90°) without Rotational Augmentation}
        \label{fig:rotaug_output_norot_90}
    \end{subfigure}
    \begin{subfigure}{0.22\textwidth}
        \centering
        \includegraphics[width=\linewidth]{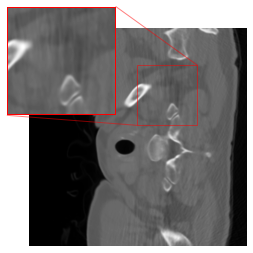}
        \caption{Output (0°) with Rotational Augmentation}
        \label{fig:rotaug_output_rot_0}
    \end{subfigure}
    \begin{subfigure}{0.22\textwidth}
        \centering
        \includegraphics[width=\linewidth]{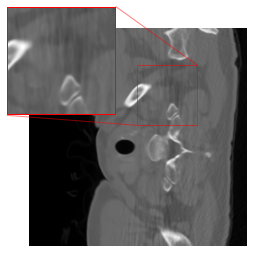}
        \caption{Output (90°) with Rotational Augmentation}
        \label{fig:rotaug_output_rot_90}
    \end{subfigure}
    
    \caption{Comparison of network performance with and without rotational augmentation on the LoDoPaB-CT test dataset. The outputs for the network with rotational augmentation remain consistent (e, f), regardless of the input image's orientation (0° and 90°). In contrast, the network without rotational augmentation shows variation in output between normal and rotated input scenarios (c, d). 
    (The rotated versions are presented in an unrotated fashion for easier comparison.)}
    \label{fig:rot_augmentation}
\end{figure}

To evaluate the impact of rotational augmentation, we conducted an experiment by training two networks: one with rotational augmentation and the other without. The objective was to investigate the network's response to input images under different orientations. Specifically, we applied two scenarios:
firstly, inputting images as they are, and secondly, rotating them by 90° before feeding them into the networks.
Results depicted in Fig. \ref{fig:rot_augmentation} illustrate a noteworthy observation. For the network with rotational augmentation during training (cf. \ref{fig:rotaug_output_rot_0} and \ref{fig:rotaug_output_rot_90}), the output remains consistent regardless of the input image's orientation. However, for the network trained without rotational augmentation (cf. \ref{fig:rotaug_output_norot_0} and \ref{fig:rotaug_output_norot_90}), the output varies between the normal and rotated input scenarios.
This experiment highlights the effectiveness of rotational augmentation in achieving a robust network 
and ensuring consistent performance across different orientations of input images.

\subsection{Effectiveness of Gaussian Noise Augmentation}

In continuation of our augmentation strategy, we introduce an experiment aimed at evaluating the influence of Gaussian noise augmentation on network performance. 
The experiment involved a comparative analysis of three scenarios:
(i) Using the dataset without Gaussian noise augmentation.
(ii) Employing the entire dataset in two ways: one without noise and one with 1\% Gaussian noise.
(iii) Utilizing the entire dataset in three ways: one without noise, one with 0.5\% Gaussian noise, and one with 1\% Gaussian noise.
The results based on PSNR and SSIM, presented in Table \ref{tab:gauss_aug_comparison}, reveal that the methodology employed in Experiment (ii) exhibited slightly better
performance compared to the other two cases.

\setcounter{table}{4}

\begin{table}[h]
    \centering
    \begin{tabular}{lcc}
        \toprule
        & \textbf{PSNR} & \textbf{SSIM} \\
        \midrule
        \textbf{Experiment (i)} & 35.88 $\pm$ 4.709 & 0.8463 $\pm$ 0.1515 \\
        \textbf{Experiment (ii)} & 35.91 $\pm$ 4.750 & 0.8466 $\pm$ 0.1513 \\
        \textbf{Experiment (iii)} & 35.86 $\pm$ 4.767 & 0.8457 $\pm$ 0.1520 \\
        \bottomrule
    \end{tabular}
    \caption{Performance comparison of PSNR and SSIM metrics on the LoDoPaB-CT test dataset between 1) using the dataset without Gaussian noise augmentation, 2) employing the entire dataset in two ways: one without noise and one with 1\% Gaussian noise and 3) utilizing the entire dataset in three ways: one without noise, one with 0.5\% Gaussian noise, and one with 1\% Gaussian noise.}
    \label{tab:gauss_aug_comparison}
\end{table}

\subsection{Impact of the Filter Type for the FBP Preprocessing}

Another important aspect of the implementation involves the selection of the filter function employed in the FBP reconstruction process. 
We conducted experiments using two well-known filter types: the Hann filter and the Ram-Lak filter. 
The evaluation results are presented in Table \ref{tab:hann_ramlak_comparison}. 
Analysis of the outcomes leads to the conclusion that the Ram-Lak filter is the preferable choice for this specific configuration.
This preference is attributed to the behavior of the filters; the Hann filter, by attenuating high frequencies, induces greater blurriness in the resulting FBP compared to the Ram-Lak filter, which solely considers the absolute value. 
The reconstruction of a slightly blurry image poses a more challenging task for the DRUNet.

\begin{table}
    \centering
    \begin{tabular}{lcc}
        \toprule
        & \textbf{PSNR} & \textbf{SSIM} \\
        \midrule
        \textbf{DRUNet with FBP using} & $36.13  \pm 4.901$ & $0.8514 \pm 0.1471$ \\
        \textbf{Hann filter} & & \\
        \textbf{DRUNet with FBP using} & $36.23 \pm 4.894$ & $0.8528 \pm 0.1463$ \\
        \textbf{Ram-Lak filter} & & \\
        \bottomrule
    \end{tabular}
    \caption{Performance comparison of PSNR and SSIM metrics on the LoDoPaB-CT test dataset between the DRUNet trained on FBP reconstructions using the Hann filter and the Ram-Lak filter.}
    \label{tab:hann_ramlak_comparison}
\end{table}

\subsection{Hyperparameter Tuning for the SSIM-based Loss Function}

To optimize the performance of our model, we turn our attention to fine-tuning the parameter $\alpha$ within the SSIM-based loss function. 
The parameter $\alpha$ regulates the trade-off between SSIM and MAE during the loss computation. 
Based on empirical values from unrecorded previous tests, we identified $\alpha = 5$ as a good choice.
After various adjustments, we wanted to verify the parameter value again with a small ablation study.
To save computation time for the parameter verification, we conducted preliminary experiments on a subset of the LoDoPaB-CT training dataset, comprising 640 images out of the total 35820, training for 400 epochs. 
The results for different values of $\alpha$ are presented in Table \ref{tab:ssim_alpha_tuning_small_scale}. 
The results emphasize that there is minimal variation across different positive $\alpha$ values.
To emphasize the importance of balancing SSIM and MAE, we also set $\alpha$ to 0, relying solely on MAE loss with $L_2$ regularization. 
This approach resulted in a notable decline in performance.
Conversely, a higher $\alpha$ values gradually results in reduced model performance.

\begin{table}
    \centering
    \begin{tabular}{lccc}
        \toprule
        $\alpha$ & \textbf{PSNR} & \textbf{SSIM} \\
        \midrule
        0 & $35.44 \pm 4.528$ & $0.8347 \pm 0.1583$ \\
        1 & $35.94 \pm 4.729$ & $0.8470 \pm 0.1511$ \\
        3 & $35.94 \pm 4.750$ & $0.8471 \pm 0.1510$ \\
        5 & $35.91 \pm 4.750$ & $0.8466 \pm 0.1513$ \\
        7 & $35.88 \pm 4.748$ & $0.8462 \pm 0.1515$ \\
        10 & $35.86 \pm 4.747$ & $0.8460 \pm 0.1516$ \\
        \bottomrule
    \end{tabular}
    \caption{Results of tuning $\alpha$ on a small scale experiment using 640 images from the LoDoPaB-CT training dataset, with evaluation conducted on the entire LoDoPaB-CT test dataset.}
    \label{tab:ssim_alpha_tuning_small_scale}
\end{table}

To further investigate this hyperparameter choice, we performed an additional experiment on the complete LoDoPaB-CT dataset. We compared our initial setting of $\alpha = 5$ to the best performing value observed in the small scale experiment, $\alpha = 3$. The outcomes are presented in Table \ref{tab:ssim_alpha_tuning_full_scale}. The results reveal that in this specific experiment $\alpha = 5$ performs better than $\alpha = 3$, suggesting an enhanced overall reconstruction quality. Given these results, we have chosen to use $\alpha = 5$ as the preferred parameter setting for our SSIM-based loss function.

\begin{table}[h]
    \centering
    \begin{tabular}{lccc}
        \toprule
        $\alpha$ & \textbf{PSNR} & \textbf{SSIM} \\
        \midrule
        3 & $36.18 \pm 4.858$ & $0.8518 \pm 0.1470$ \\
        5 & $36.23 \pm 4.894$ & $0.8528 \pm 0.1463$ \\
        \bottomrule
    \end{tabular}
    \caption{Results of tuning $\alpha$ on the entire LoDoPaB-CT training dataset, confirming that $\alpha = 5$ provides better performance.}
    \label{tab:ssim_alpha_tuning_full_scale}
\end{table}

% Bibliography
\bibliographystyle{amsplain}
\bibliography{bare_adv.bib}

\end{document}